\pdfoutput=1
\documentclass[twocolumn]{aastex62}

\usepackage[normalem]{ulem} 

\shorttitle{Variable PMS stars in Wd2}
\shortauthors{Sabbi et al.}

\begin{document}

\title{Time-domain study of the young massive cluster Westerlund~2 with the Hubble Space Telescope. I.\footnote{Based on observations with the NASA/ESA Hubble Space Telescope, obtained at the Space Telescope Science Institute, which is operated by AURA Inc., under NASA contract NAS 5-26555.}}
\correspondingauthor{Elena Sabbi}
\email{sabbi@stsci.edu}

\author[0000-0003-2954-7643]{E. Sabbi}
\affil{Space Telescope Science Institute \\
3700 San Martin Dr \\
Baltimore, MD 21218, USA}

\author{M. Gennaro}
\affiliation{Space Telescope Science Institute \\
3700 San Martin Dr \\
Baltimore, MD 21218, USA}

\author{J. Anderson}
\affiliation{Space Telescope Science Institute \\
3700 San Martin Dr \\
Baltimore, MD 21218, USA}

\author{V. Bajaj}
\affil{Space Telescope Science Institute \\
3700 San Martin Dr \\
Baltimore, MD 21218, USA}

\author{N. Bastian}
\affiliation{Astrophysics Research Institute, Liverpool John Moores University\\
146 Brownlow Hill\\
Liverpool L3 5RF, UK}

\author{J.S. Gallagher, III}
\affiliation{Department of Astronomy, University of Wisconsin-Madison \\ 
5534 Sterling, 475 North Charter Street \\
Madison WI 53706, USA}

\author{M. Gieles}
\affiliation{Department of Physics, University of Surrey \\
Guildford \\
GU2 7XH, UK}

\author{D.J. Lennon}
\affiliation{Instituto de Astrofísica de Canarias \\
E-38200 La Laguna \\
Tenerife, Spain}
\affiliation{Departamento de Astrofísica, Universidad de La Laguna \\
E-38205 La Laguna \\
Tenerife, Spain.}

\author{A. Nota}
\affiliation{Space Telescope Science Institute \\
3700 San Martin Dr \\
Baltimore, MD 21218, USA}
\affiliation{ESA, SRE Operations Devision \\
Spain}

\author{K.C. Sahu}
\affiliation{Space Telescope Science Institute \\
3700 San Martin Dr \\
Baltimore, MD 21218, USA}

\author{P. Zeidler}
\affiliation{Space Telescope Science Institute \\
3700 San Martin Dr \\
Baltimore, MD 21218, USA}
\affiliation{Department of Physics and Astronomy, Johns Hopkins University \\  Baltimore, MD 21218, USA}


\begin{abstract}

Time-domain studies of pre-main sequence stars have long been used to investigate star properties during their early evolutionary phases and  to trace the evolution of circumstellar environments. Historically these studies have been confined to the nearest, low-density, star forming regions. We used the Wide Field Camera 3 on board of the Hubble Space Telescope  to extend, for the first time, the study of pre-main sequence variability to one of the few young massive clusters in the Milky Way, Westerlund 2. Our analysis reveals that at least 1/3 of the intermediate and low-mass pre-main sequence stars in Westerlund 2 are variable. Based on the characteristics of their light curves, we classified $\sim 11$\% of the variable stars as weak-line T-Tauri candidates, $\sim 52$\% as classical T-Tauri candidates, $\sim 5$\% as dippers and $\sim 26$\% as bursters. In addition, we found that $2\%$ of the stars below $6\,  M_\odot$ ($\sim 6$\% of the variables) are eclipsing binaries, with orbital periods shorter than 80 days. The spatial distribution of the different populations of variable pre-main sequence stars suggests that stellar feedback and UV-radiation from massive stars play an important role on the evolution of circumstellar and planetary disks.

\end{abstract}


\section{Introduction}

\begin{figure*}[ht!]
    \centering
    \includegraphics[width=\textwidth]{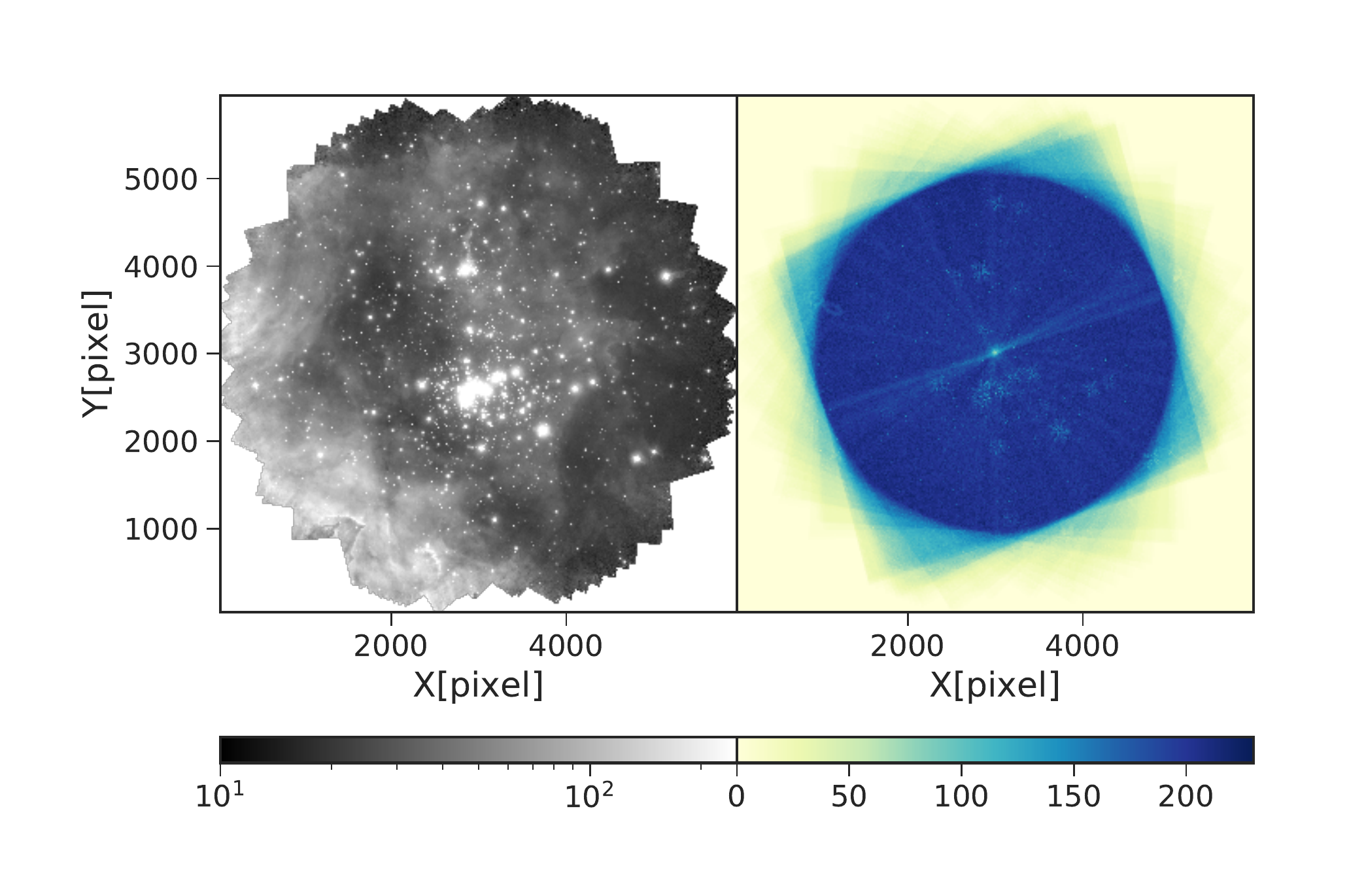}
    \caption{{\it Left Panel} -- Stack image of Wd2, obtained combining all the deep exposures acquired in the UVIS filter F814W. North is up and East to the left. The pixel scale is 40 mas/pixels. At the distance of Wd2 \citep[4.16 kpc]{Zeidler2015} this translate to 0.02 pc/pixels. {\it Right Panel} -- Coverage map of Wd2 mosaic. The darker the image, the greater is the number of overlapping exposures.}
    \label{fig:pointings}
\end{figure*}

Pre-main sequence (PMS) stars have been long known to be variable objects \citep{Joy1945}. Over the past 50 years, studies of their variability have been used, for example, to characterize how stars accrete mass from their circumstellar disks \citep[i.e.][]{Hartmann1993,Bell1994,DAngelo2010,Cody2017}, estimate PMS star rotational velocities  \citep[i.e.]{Hartmann1989, Marilli2007, Venuti2017}, infer properties of the circumstellar disks, and trace the evolution of the circumstellar environment \citep[e.g.][]{Hillenbrand2002}. 

Young star clusters are excellent laboratories to study the evolution of PMS stars and their circumstellar disks, as they provide rich samples of objects having approximately the same age, distance and chemical composition over a wide range of masses. These systems are characterized by large dust extinction, bright background level, and high stellar crowding. Therefore the study of PMS star time series have been, so far, limited to the closer and/or lower stellar density star forming regions, such as the Orion Nebula Cluster \citep[e.g.][]{Stassun1999, Herbst2002, Stassun2007, Rice2015}, NGC~1893 \citep{Lata2012}, $\rho$~Oph and Upper~Sco \citep{Cody2017}, NGC~2264 \citep{Alencar2010, Cody2017, Cody2018} and Stock~8 \citep{Lata2019}, thus covering only a limited range of parameters in terms of star formation rate, stellar density and feedback.

The exquisite sensitivity and spatial resolution of the Hubble Space Telescope (HST) allow us to extend time-domain studies to the population of PMS stars in young clusters (age $< 5\, {\rm Myr}$) more massive than $10^4\, \rm{M_\odot}$ (a.k.a. young massive clusters, YMCs). These systems trace intense episodes of star formation (SF), and because they are bright, can be observed at several tens of Mpc from us, in interacting and starburst galaxies. 

\begin{figure*}[ht!]
\centering
\includegraphics[width=\textwidth]{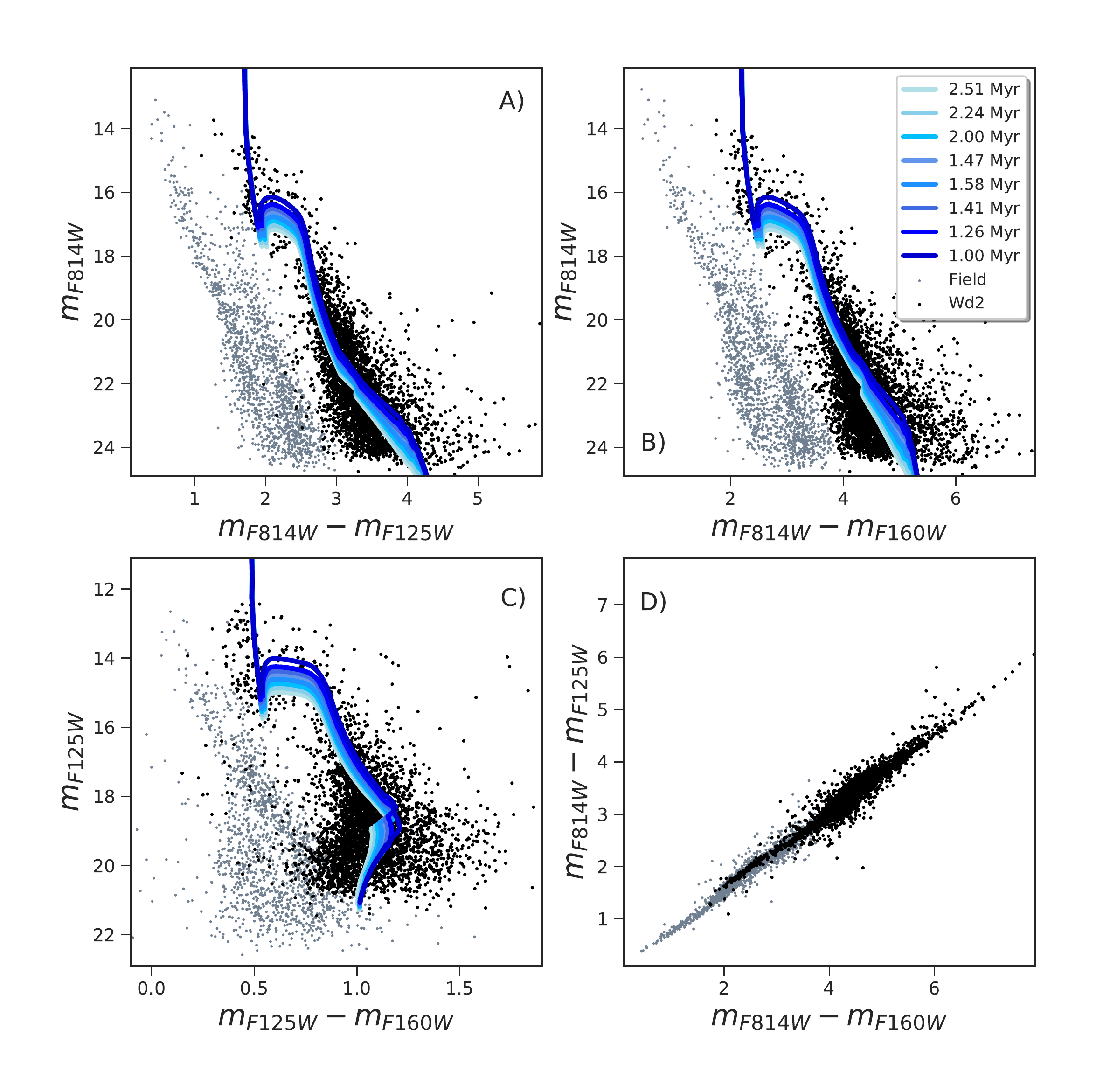}
\caption{Wd2 color-magnitude diagrams for the filter combinations $m_{F814W}$ vs $m_{F814W}-m_{F125W}$ ({\it Panel A)}), $m_{F814}$ vs $m_{F814W}-m_{F160W}$ ({\it Panel B)}), and $m_{F814}$ vs $m_{F814W}-m_{F125W}$ ({\it Panel C)}). {\it Panel D)} shows the color-color diagram $m_{F814}-m_{F160W}$ vs $m_{F814W}-m_{F125W}$. Cluster candidates are shown as black dots, while gray points are used for the stars in the field of the MW. Padova isochrones for solar metallicity and ages between 1.0 and 2.5 Myr are plotted over the cluster stellar population for reference, by assuming the distance of 4.16 kpc and the average extinction $R_v=3.8$, that we derived in \citet{Zeidler2015}. }
\label{fig:cmds}
\end{figure*}

\begin{figure*}[ht!]
\centering
\includegraphics[width=\textwidth]{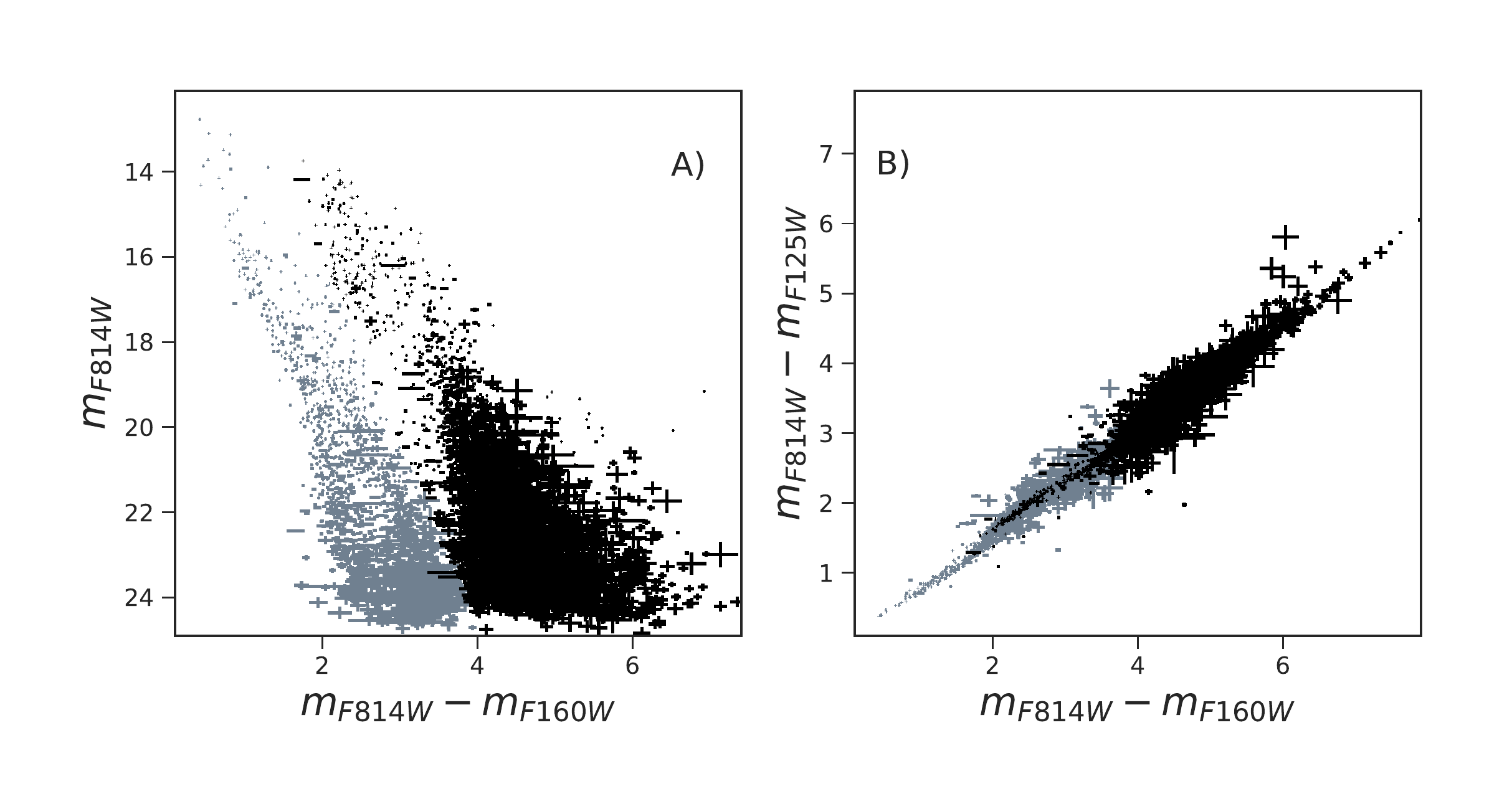}
\caption{{\it Panel A)} CMD $m_{F814}$ vs $m_{F814W}-m_{F160W}$. The size of the symbols represents the photometric errors. {\it Panel D)} The color-color diagram $m_{F814}-m_{F160W}$ vs $m_{F814W}-m_{F125W}$. As in Panel A) the symbol size measures the photometric error.}
\label{fig:phot_ers}
\end{figure*}

In the Milky Way (MW) and in the Local Group (LG) YMCs are rare. Yet these few examples, close enough to be resolved into stars, are of prime interest for studying the formation and evolution of stars over more than three order of magnitudes (from over $100\, M_\odot$ down to the hydrogen burning limit $\sim 0.1\, M_\odot$) in regions characterized by extreme stellar densities and UV radiation. Time-series analysis of nearby YMCs, thus, offer the unique opportunity to collect rich samples of variable stars over a wide range of masses to characterize the earlier phases of PMS star evolution, mass accretion rates, and circumstellar disk survivability in environments that, in many ways, resemble the extreme conditions found in more distant interacting and starburst galaxies, in the early phases of globular cluster formation and in the early Universe. 

In this paper we present the first HST time-domain study of the galactic YMC Westerlund 2 \citep[Wd2;][]{Westerlund1961}. With an age of one to two Myr \citep{Vargas2013, Zeidler2015} and a stellar mass of $M=3.6\times 10^4\, {\rm M_\odot}$ \citep{Zeidler2017}, Wd2 is one of the most massive YMCs in the MW. The cluster is located in the Sagittarius spiral arm, at $\simeq4.16\, {\rm kpc}$ from the Sun \citep{Vargas2013, Zeidler2015}. It hosts a rich population of OB-type stars \citep{Moffat1991, Vargas2013}, including a double-line binary WR-star with component minimum masses of $83.0\pm 5.0\, {\rm M_\odot}$ and $82.0\pm 5.0\, {\rm M_\odot}$ respectively \citep{Bonanos2004}. 

Using the HST Advanced Camera for Surveys (ACS; filters F555W, F658N and F814W) and the Wide Field Camera (WFC3) IR channel (filters F125W, F128N, and F160W, PI Nota, GO-13038) we found that the cluster consists of two nearly coeval clumps \citep{Zeidler2015}, and that is mass segregated \citep{Zeidler2017}. Following the same approach described in \citet{DeMarchi2010}, we identified a rich population of PMS stars with H$\alpha$ excess, that are likely still accreting material from their circumestellar disks \citep{Zeidler2016}

The paper is organized as follows. In Section~\ref{Obs} we present the data, describe how we created the reference frame and performed the photometric reduction. We discuss the characteristics of the color-magnitude diagrams (CMDs) in Section~\ref{the CMDs}, and the selection criteria used to identify the variable stars in Section~\ref{Variables}. We classify the variable stars and describe their properties in Section~\ref{Variable classificiaton}. Conclusions are presented in Section~\ref{Conclusions}.  

\section{Observations and Data Reduction}
\label{Obs}

\subsection{The Data}
\label{the data}

\begin{figure*}
    \centering
    \includegraphics[width=\textwidth]{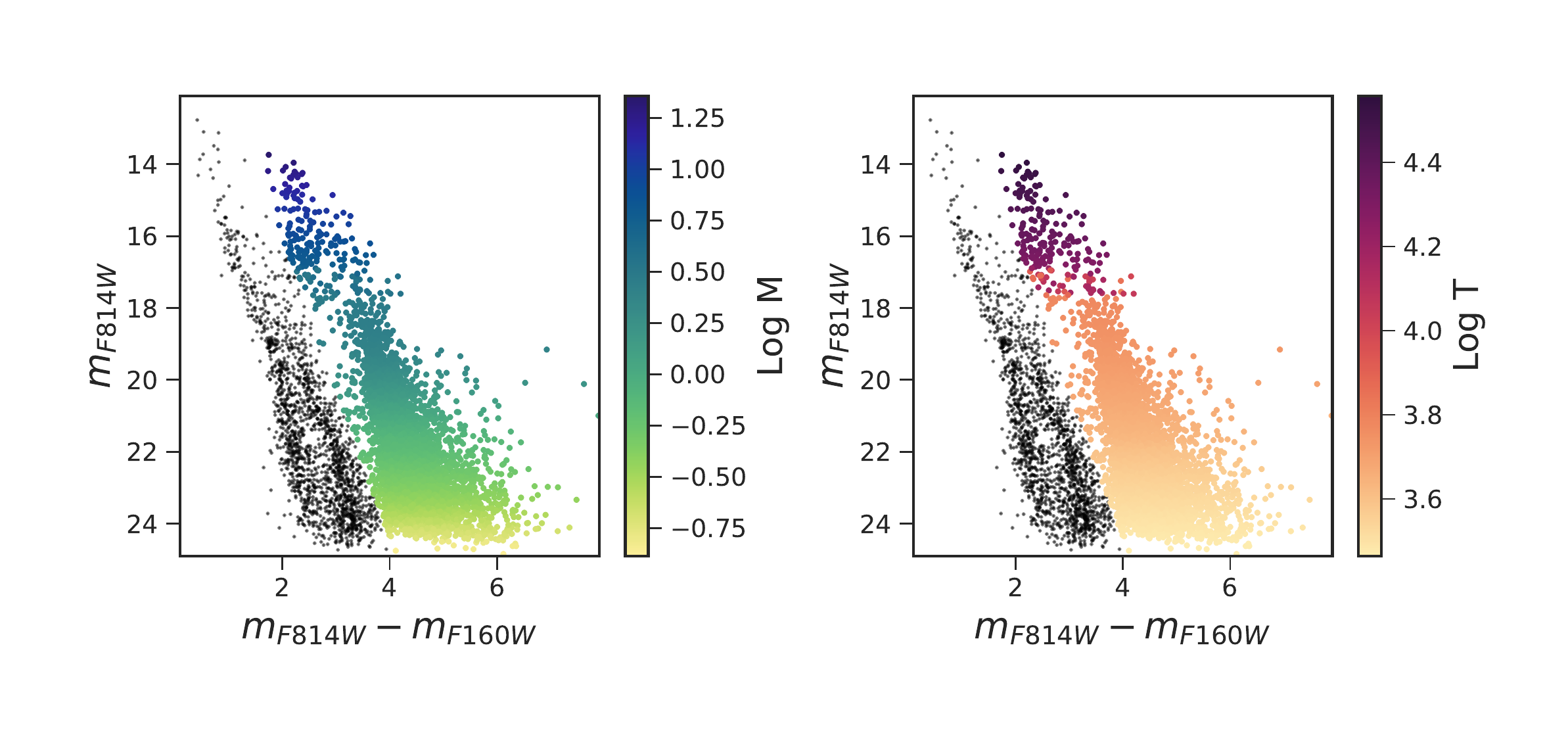}
    \caption{CMD of Wd2 for the filter combination $m_{F814}$ vs $m_{F814W}-m_{F160W}$. In the {\it Left Panel} cluster stars are color-coded based on their mass, and on the {\it Right Panel} Wd2 stars are color-coded based on their temperature. Temperatures and masses where derived using the 1.6 Myr old Padova isochrones for solar metallicity, assuming a distance of 4.16 kpc and average extinction $R_v=3.8$. In both cases we used a logarithmic scale. MW field stars are shown as black dots.}
    \label{fig:mass&temp}
\end{figure*}

We repeatedly observed the YMC Wd2 from October 2016 to October 2019 for a total of 47 orbits of HST, using the UVIS channel of WFC3 (PI Sabbi, GOs-14087, 15362, 15514). Each orbit consists of two 3 sec (hereafter short exposures) and six 350 sec  (hereafter long exposures) observations in the filter F814W. For the long exposures, we adopted a six-point dither pattern with a step of $1.26\arcsec$ to guarantee an optimal sampling of the PSF, fill the gap between the UVIS detectors, and allow for hot pixel removal. The two short exposures were acquired at the beginning and at the end of each orbit respectively, at the same position of the first and last long exposures. To identify and characterize the various types of Wd2 variable stars (periodic, semi-periodic and non-periodic), we organized the observations in groups of 5 or 10 visits using a non-uniform, logarithmic increasing spacing. 

In this paper we are limiting our analysis to the long exposures, acquired during the first 26 visits, taken between October $30^{th}$ 2016 to July $5^{th}$ 2018. Over this period we collected a total of 156 long exposures organized as follows: 
\begin{itemize}
\item The first 5 epochs were separated by 51 days each, and were acquired between October 2017 and June 2017; 
\item In August 2017 we observed Wd2 10 times with a 0.8 days cadence;
\item And 1.6 days later we obtained other 5 epochs, each separated by 1.6 days;
\item The first epoch belonging to the 102 day cadence was observed on November 27$^{th}$, 2017
\item And 5 epochs separated by 3.2 days, were acquired between March $9^{th}$, and July $5^{th}$, 2018;
\end{itemize}

In \citet{Zeidler2015} we showed that Wd2 consists of two nearly coeval clumps. To guarantee the best coverage of the entire cluster population, we centered the gap among the two UVIS CCDs  between the two clumps, in a region almost devoid of stars. To allow for a more flexible scheduling of the observations,  we did not request any specific orientation. The {\it left panel} of Fig.~\ref{fig:pointings} shows the stack image, obtained by combining all deep exposures, while the {\it right panel} shows the coverage map of the survey, and highlights how many images contributed to each pixel. 

\subsection{The Reference Catalog}
\label{The photometry}

We carried out the photometric analysis directly on the bias-subtracted, flat-fielded and charge-transfer-efficiency (CTE) corrected {\tt \_flc.fits} exposures produced by the standard calibration pipeline calwf3 v3.5\footnote{Gennaro, M., et al. 2018, ''WFC3 Data Handbook'', version 4.0 http://www.stsci.edu/hst/wfc3/documents/handbooks/currentDHB/wfc3\_dhb.pdf}. Compared to drizzled ({\tt \_drc}) images,  {\tt \_flc} images have the advantage of not being re-sampled and therefore they are the most direct representation of the astronomical scene. However {\tt \_flc} images are still affected by geometric distortion, that we took into account by creating a distortion-free reference frame, and relating the photometry and astrometry of each exposure to that frame.

\begin{figure*}[ht!]
\centering
\includegraphics[width=\textwidth]{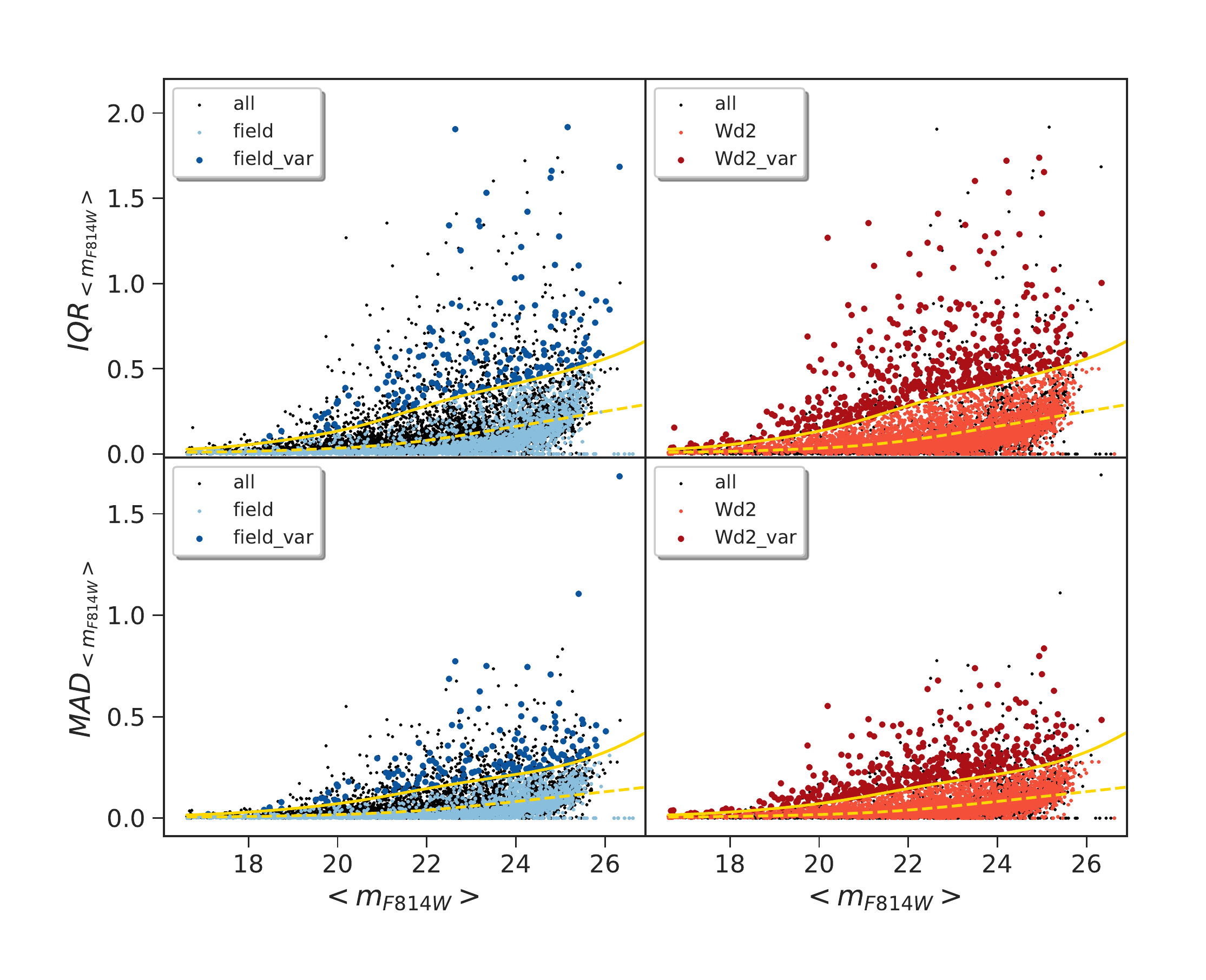}
\caption{{\it Upper Panels:} Distribution of the IQR value as a function of magnitude for the stars that have been observed in at least 16 visits. Field stars are shown in the ({\it Left Panel}) in blue and Wd2 candidates are shown in red in the  ({\it Right Panel}). The yellow dashed line marks the running mean of the entire population, the continuous yellow line is $1.349\times\sigma_{IQR}$. {\it Lower Panels:} as the {\it Upper Panels} but for the MAD statistics. The continuous yellow line represents $1.4826\times\sigma_{MAD}$. }
\label{fig:var_sel}
\end{figure*}

We used the publicly available one-pass photometry routine {\tt img2xym} \citep{Anderson2006} and a library of empirical PSFs to identify and measure all the high signal-to-noise ($S/N> 150$), isolated (within 5 pixel radius), and non-saturated sources. The PSF library properly accounts for the spatial variations caused by the optics of the telescope and the variable charge diffusion in the CCD. Seasonal and daily thermal changes, however, can cause variations in the optical path length of HST up to few microns within the time scales of an orbit \citep{Bely1993, Lallo2006}. These changes affect the focus of the telescope and translate to small, but measurable differences in the PSF from one exposure to another. To take these variations into account, for each exposure we derived the correction necessary to minimize its average PSF-residuals. We then derived for each catalog a six-parameter transformation to align the UVIS observations to the Gaia Data Release 2 \citep[DR2,][]{Lindegren2018}, and to create a distortion-free reference frame with a pixel scale of 40 mas/pixel, the UVIS native pixel scale. 

To separate candidate members of Wd2 from the field of the MW, and to estimate masses and temperatures of the stars in the cluster, we combined our data with our previous WFC3/IR observations acquired in the filters F125W and F160W (PI Nota, GO-13038). We aligned the IR observations to Gaia using the same procedure followed for UVIS.

\begin{figure*}[ht!]
\centering
\includegraphics[width=\textwidth]{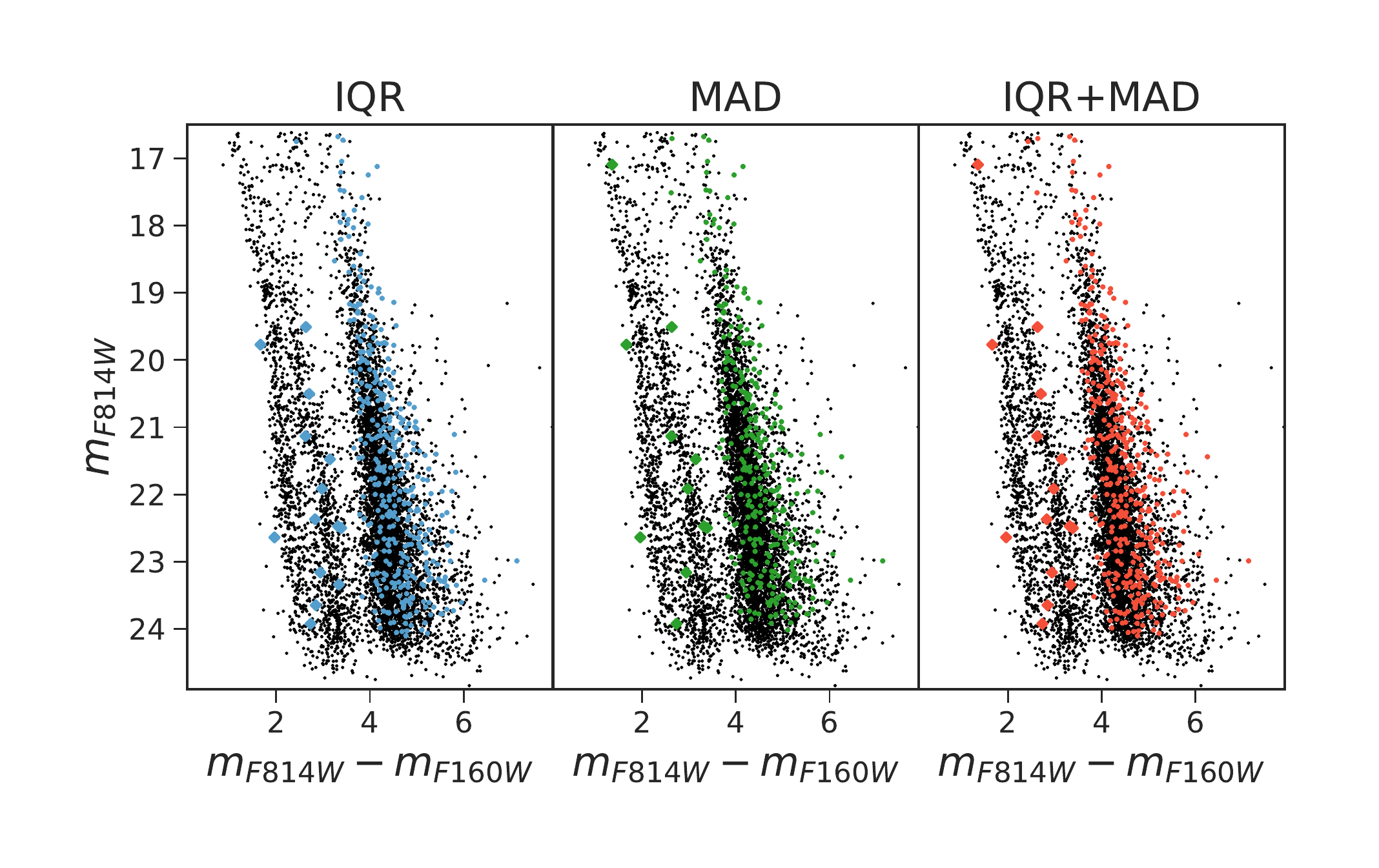}
\caption{Colors and magnitudes of the variable stars as detected using the IQR statistics (shown as blue points in {\it Left Panel)}, the MAD statistics (shown as green points in the {\it Middle Panel}) and a combination of both (red points in the {\it Right Panel}).}
\label{fig:variable_CMD}
\end{figure*}

The final photometric analysis was performed using the software KS2 \citep[Anderson, in preparation; see][ for details]{Sabbi2013,Sabbi2016}, that can use all the exposures together to detect, measure and subtract stars in subsequent passes, moving from the brightest to the fainter sources. KS2 can perform simultaneous detection in multiple filters. However, given the higher spatial resolution of the UVIS data-set, we only used the long exposures acquired through the UVIS/F814W filter to create the reference catalogds. For each source KS2 provides the mean flux, the flux standard deviation and an estimate of the quality of the fit of the PSF. The final catalogs contains in total 9,268 sources.

The fluxes of the sources found by KS2 in the UVIS/F814W combined observations were measured in the WFC3/IR exposures, and in each of the individual F814W visits. For each exposure we used its own ``focus-specific'' PSF-library, derived as described above. 

Fluxes were converted into instrumental magnitudes and then calibrated in the Vega-mag photometric system following the same approach described in \citet{Sabbi2016}, and using the zero points listed on the STScI Web site\footnote{http://www.stsci.edu/hst/instrumentation/wfc3/data-analysis/photometric-calibration/uvis-photometric-calibration}. The aperture correction was derived using a $0.\arcsec 4$ aperture photometry on the drizzled images.

\section{The Color-Magnitude Diagrams}
\label{the CMDs}

Fig.~\ref{fig:cmds} shows Wd2 color-magnitude diagrams (CMDs) and the color-color diagram in the combination of filters: 
\begin{itemize}
    \item[] {\it Panel A): } $m_{F814}$ vs $m_{F814W}-m_{F125W}$ ,
    \item[] {\it Panel B):} $m_{F814}$ vs $m_{F814W}-m_{F160W}$ , 
    \item[] {\it Panel C):} $m_{F814}$ vs $m_{F814W}-m_{F125W}$ . 
    \item[] {\it Panel D):} shows the color-color diagram $m_{F814}-m_{F160W}$ vs $m_{F814W}-m_{F125W}$. 
\end{itemize} 
From now on in each plot we show mean magnitudes and colors as derived from the combination of all the single exposures.

\begin{figure*}[ht!]
\centering
\includegraphics[width=15.cm]{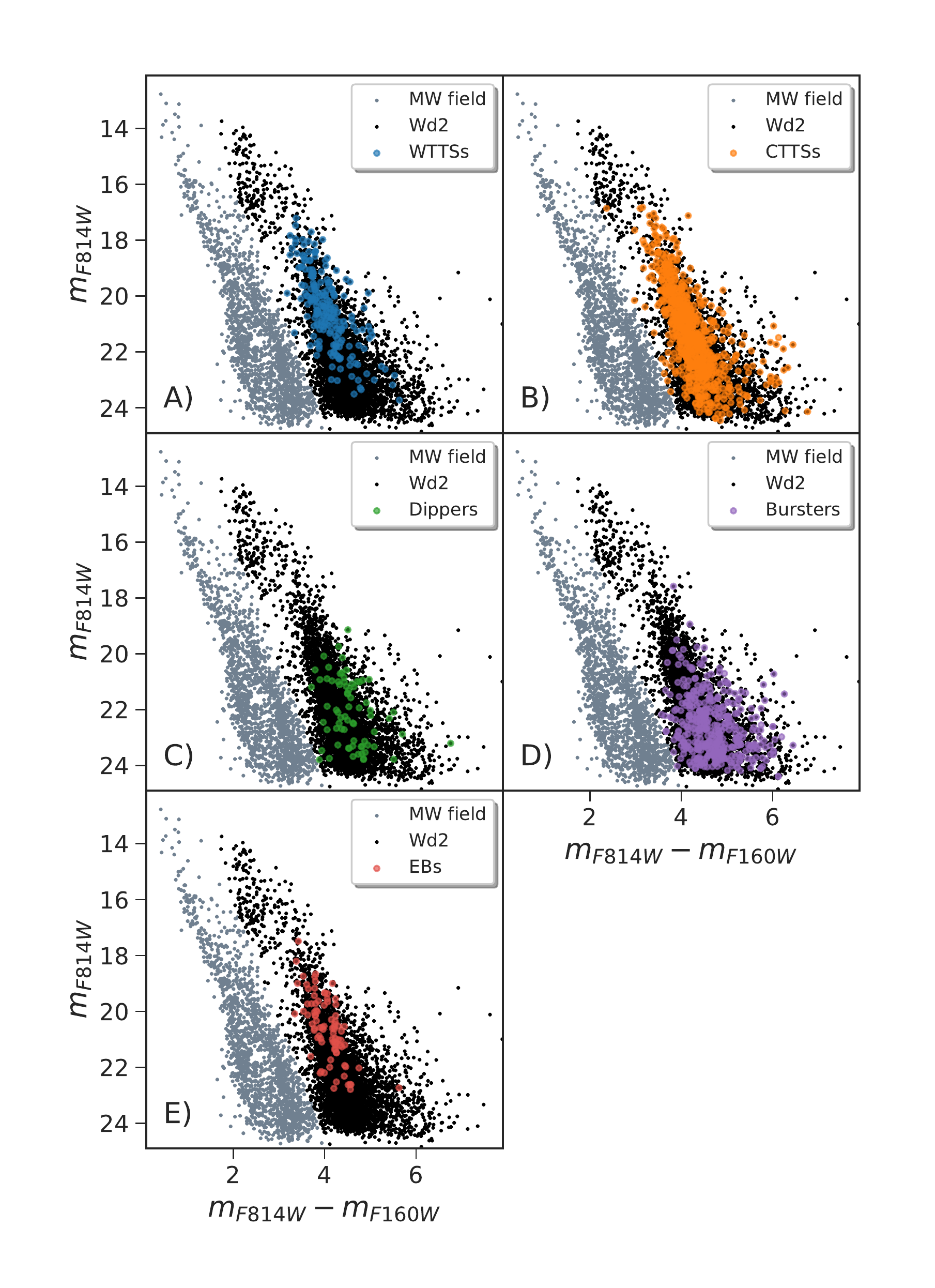}
\caption{Colors and magnitudes of the five types of variable stars. WTTS candidates are shown in blue in {\it Panel A)}. CTTS candidates are shown in orange in {\it Panel B)}. {\it Panel C)} shows colors and magnitudes of dipper candidates as green points, while burster candidates are shown in {\it Panel D)} as purple points. Candidates eclipsing binaries are shown in {\it Panel E)} in red.}
\label{fig:variable_CMD2}
\end{figure*}

\begin{figure*}[ht!]
\centering
\includegraphics[width=15.cm]{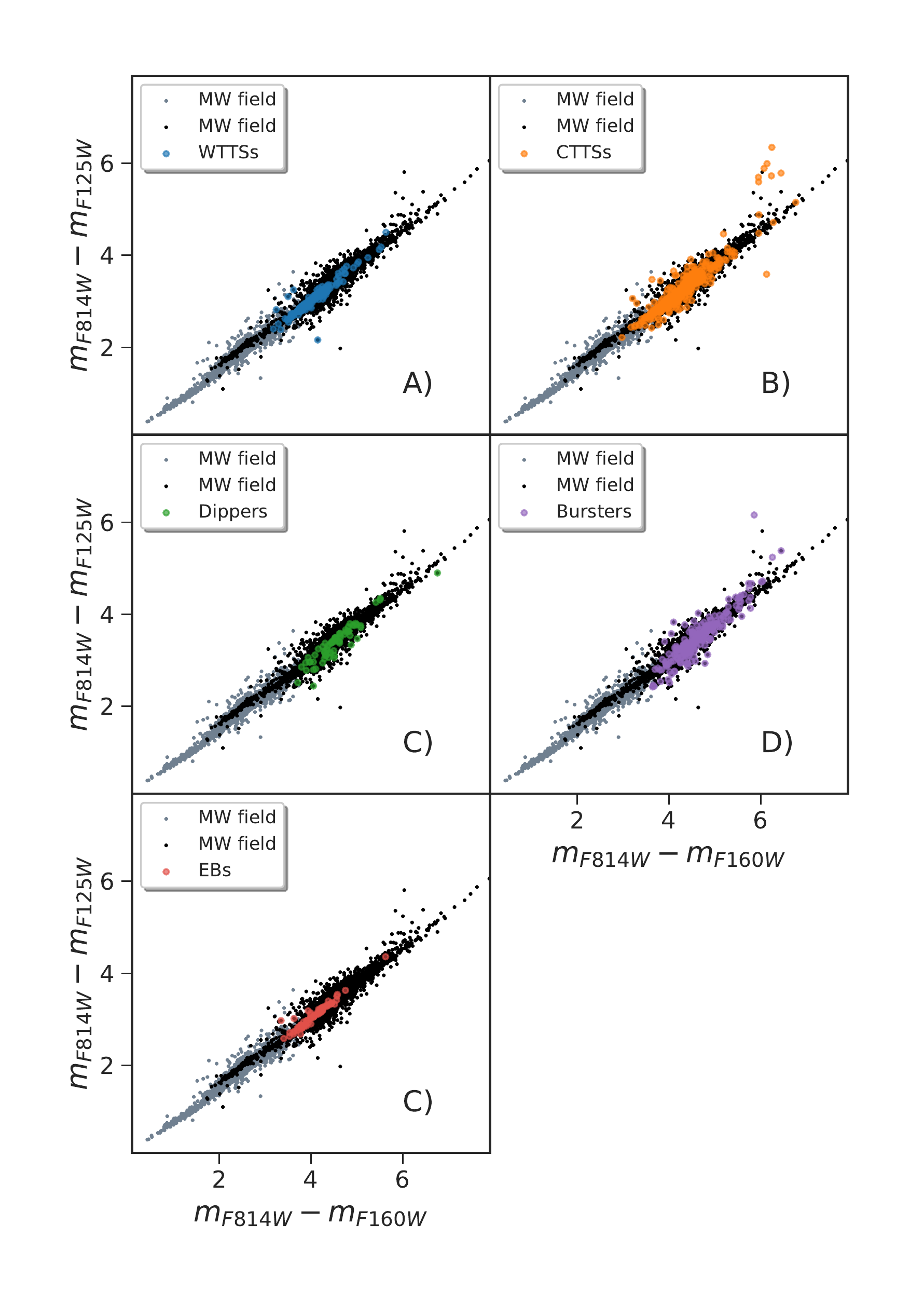}
\caption{Wd2 color-color diagram with the five classes of variables superimposed using the same color-scheme adopted in Fig.~\ref{fig:variable_CMD2}.}
\label{fig:variable_CC}
\end{figure*}

In all three CMDs shown in Fig.~\ref{fig:cmds} it is possible to identify two well separated populations, with the bluer sequence belonging to the field of the MW. Wd2 intermediate and high-mass main sequence (MS) stars ($m_{F814W}\le 15.7$) and low-mass PMS stars ($m_{F814W}\ge 17.5$) stars occupy the redder sequence. The Turn On (TOn), the point were the PMS stars arrive in MSs, is well defined between $15.7<m_{F814W}<17.5$

\begin{figure*}[ht!]
\centering
\includegraphics[width=15.cm]{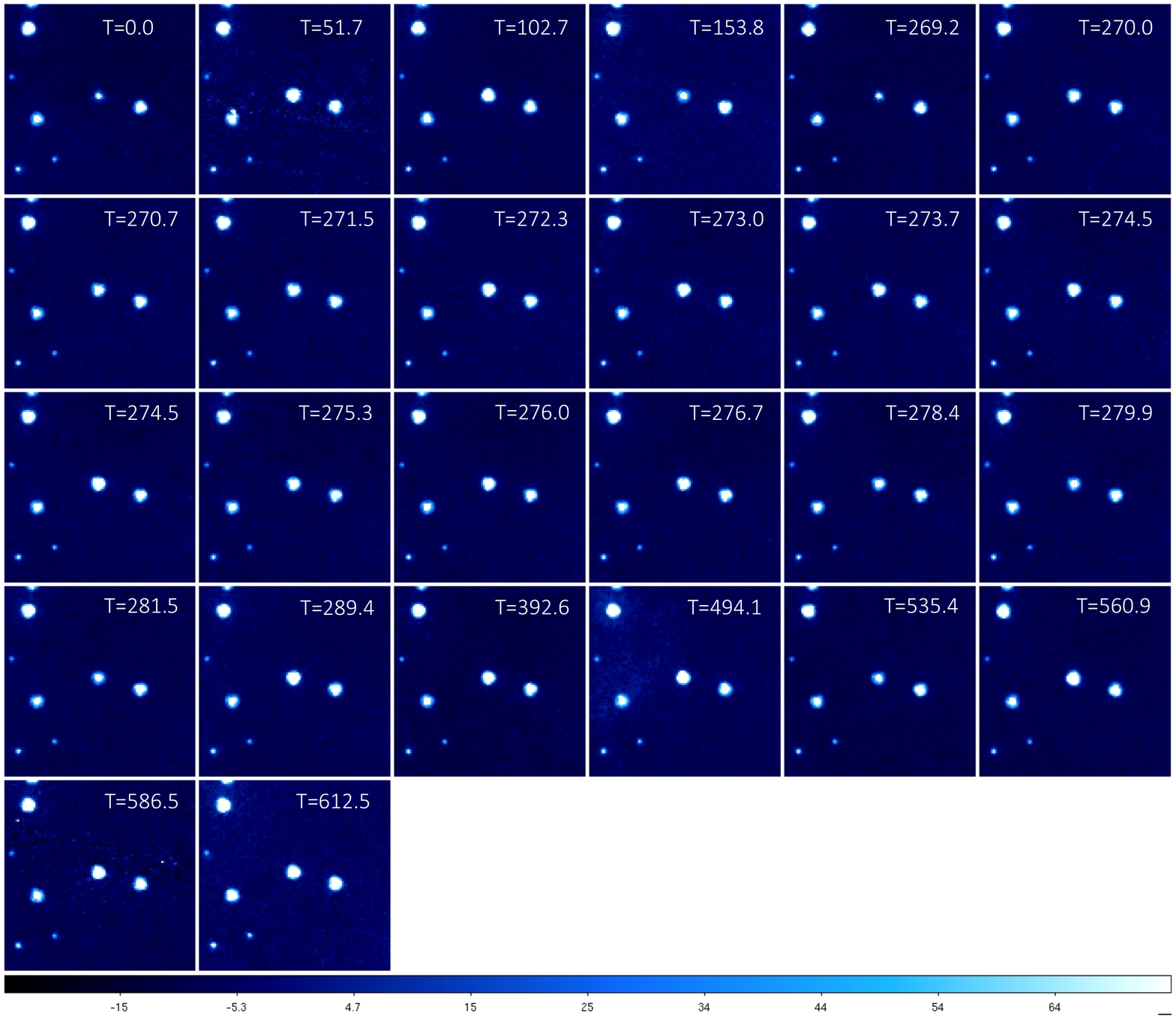}
\caption{Images of star $\# 5320$ acquired between October 2016 and July 2018 in the filter F814W. North is up and East is left. On October $30^{th}$ 2018 the star was at its minimum luminosity, corresponding to $m_{F814W}=23.26$. In 51 days its magnitude rose to $m_{F814W}=21.423$ and remained roughly constant for the next 100 days. One hundred days later the magnitude dropped again to $m_{F814W}=23.16$. After that, the star maintained an average magnitude $m_{F814W}=21.69\pm 0.32$, with an average photometric error of 0.016.
Time in days from the first observation is shown at top of each image.}
\label{fig:mag_change}
\end{figure*}

We defined as candidate members of Wd2, the 6,133 sources whose mean magnitudes met the selection criteria $6.7\times(m_{F814W}-m_{F160W})-1.59\times(m_{F814W}-17.1)>0$. In all four plots of Fig.~\ref{fig:cmds}, cluster candidates are shown as black dots, while field stars are in gray. Padova isochrones \citep{Marigo2017, Pastorelli2019} for solar metallicity and in the age range between 1.0 and 2.5 Myr are shown for reference, assuming the same distance $D=4.16$ kpc and average extinction $R_v=3.8$, that we derived in \citet{Zeidler2015}.

In the F814W filter the photometric errors, determined using the formula $\sigma_{mag}=1.1\sigma_{flux}/flux$, range from 0.001 to 0.787, and in the filter F160W from 0.001 to 0.866. The effect of the photometric errors at the various magnitudes and colors of the $m_{F814}$ vs $m_{F814W}-m_{F160W}$  CMD are shown in {\it Panel A)} of Fig.~\ref{fig:phot_ers}. Down to magnitude $m_{F814W}\simeq 23$ the photometric errors are small enough to not affect the separation between cluster and MW field. The effect of the photometric errors in the color-color diagram are shown in {\it Panel B)} of the figure.

We used the 1.58 Myr isochrone to infer an estimate of the masses and temperatures of the sources that likely belong to the Wd2 cluster, and we found that in our observations we cover the mass range between 0.1 and 22.7 $M_\odot$ and temperatures between 3,000 and 36,000 K. We emphasize that these values are highly model dependent, and should be taken only as an indication of the properties of the stellar population, and not as the definitive value of a specific source. The distribution of masses and temperature in the $m_{F814}$ vs $m_{F814W}-m_{F160W}$ CMD is shown in Fig. \ref{fig:mass&temp} using a logarithmic scale. In this paper we limit our analysis to the 4,950 sources fainter than $m_{F814W}\la 16.62$ that likely belong to the Wd2 population. Using Padova iscochrones, this corresponds to the mass range between $0.1\la M \la 5.8\, {\rm M_\odot}$. 

\section{Variability Search}
\label{Variables}

\begin{figure}[ht!]
\center
\includegraphics[width=8.4cm]{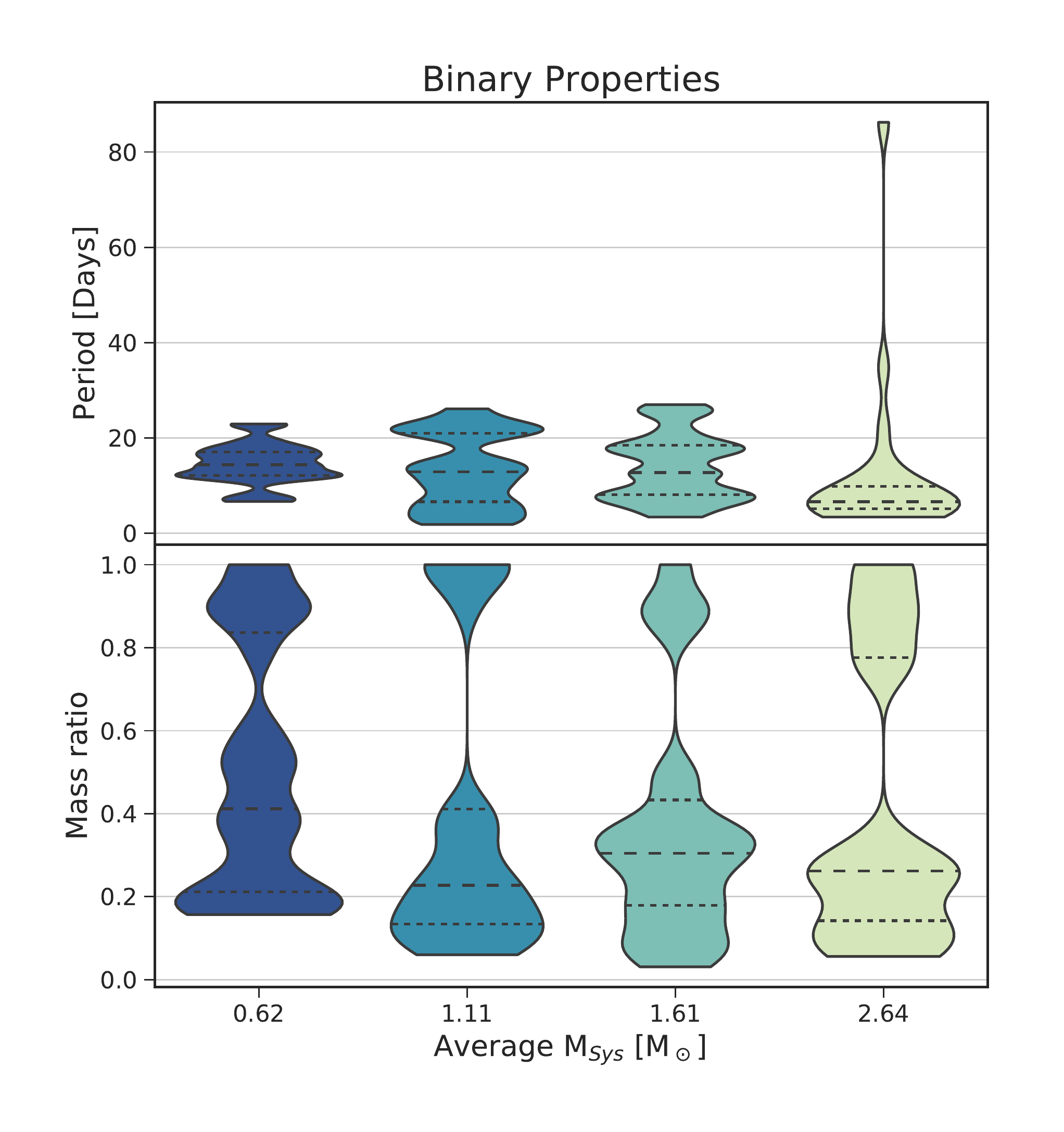}
\caption{Violin plots for binary systems in four different mass bins. The {\it Upper Panel} shows the distribution of periods, expressed in days, while the {\it Lower Panel} show the distribution of mass ratios Q. In both panels the dashed line marks the median value of the distribution, and the two dotted lines show the first and third quartile respectively. The width of each violin is scaled by the number of data in that bin.
}
\label{fig:EB_properties}
\end{figure}

The {\it Right Panel} of Fig.~\ref{fig:pointings} shows that the innermost $80\arcsec$ from the center of the image are uniformly covered by the observations, while between 80 and $114\arcsec$ the number of exposures per pixel is considerably variable. To guarantee sufficient statistics, we limited the search for variability to those sources that were observed in at least 16 different visits. 

To separate true variable sources from artifacts caused by cosmetic defects of the UVIS CCDs, diffraction spikes from saturated neighbours, and cosmic ray hits, we used two variability indices: the median absolute deviation \citep[MAD,][]{Rousseeuw1993,Richards2011}, and the interquartile range \citep[IQR,][]{Kim2014}.

The MAD index measures the scatter of observations $m_i$ and is defined as MAD$=$median$(|$m$_i-$median$($m$_i)|)$. The MAD statistic is largely used because it is insensitive to outliers \citep{Zhang2016}, however this makes it also insensitive to real occasional variations, such as the Algol-type eclipses \citep{Sokolovsky2017}. The IQR index is defined as the difference between the median values computed for the upper and lower halves of a data set, and it is preferable to MAD when dealing with an asymmetric distribution, and to detect the signal coming from eclipsing binaries. 

PMS stars are known to be variable sources \citep{Joy1945}. On the contrary we do not expect to find a high number of variable stars in the field of the MW, mainly populated by MS stars. Fig.~\ref{fig:var_sel} shows the IQR ({\it Upper Panels}) and MAD ({\it Lower Panels}) statistics for our entire sample as a function of magnitude. The dashed yellow lines represents the mean of the distributions, while the continuous yellow lines mark the $1.349\times\sigma_{IQR}$, and $1.4826\times\sigma_{MAD}$ values respectively. MW stars are shown in blue in the {\it Left Panels}, with variable candidates marked with larger dark-blue dots. Similarly, Wd2 stars are shown in orange in the {\it Right Panels}, and Wd2 variable candidates are shown as larger dark-red dots. The corresponding position of the variable stars on the $m_{F814}$ vs $m_{F814W}-m_{F160W}$ CMD is shown in Fig.~\ref{fig:variable_CMD}. 

\section{Variables Classification}
\label{Variable classificiaton}

At least 30\% (1473) of Wd2 stars between $0.1\la M \la 5.8$ $M_\odot$ are variable, with peak-to-peak magnitude variations in the F814W filter ranging from 0.035 to 3.224 magnitudes. We divided our sample in variables with peak-to-peak variations $\Delta m_{F814W}$ below and above 0.75 magnitude. We used the Lomb-Scargle  \citep{Lomb1976, Scargle1982} periodogram, a well-known algorithm for detecting and characterizing periodic signals in unevenly-sampled data, to find for each source its most plausible period (if any), and then to phase-fold its light curve (LC). 

We inspected each LC by eye, and divided our sample in five groups:
\begin{itemize}
    \item[] 168 periodic sources with $\Delta m_{F814W}\le 0.75$ were classified as weak-line T Tauri star (WTTS) candidates (these sources are further described in Section~\ref{wtts});
    \item[] 757 non-periodic sources with $\Delta m_{F814W}\le 0.75$ were classified as classical T Tauri star (CTTS) candidates (see Section~\ref{ctts} for more details); 
    \item[] 72 non-periodic sources that showed decrease in magnitude $\Delta m_{F814W}\ge 0.75$ were classified as dippers (these sources are discussed in Section~\ref{dippers});
    \item[] 379 non-periodic sources that showed outbursts in magnitude $\Delta m_{F814W}\ge 0.75$ were classified as bursters (the characteristics of bursters are presented in Section~\ref{bursters});
    \item[] 87 periodic sources with $\Delta m_{F814W}\le 0.75$ were classified as eclipsing binary (EB) candidates (further discussion about EBs is presented in Section~\ref{ebs}).
\end{itemize}

\subsection{WTTS candidates}
\label{wtts}

\begin{figure*}[ht!]
\centering
\includegraphics[width=15.cm]{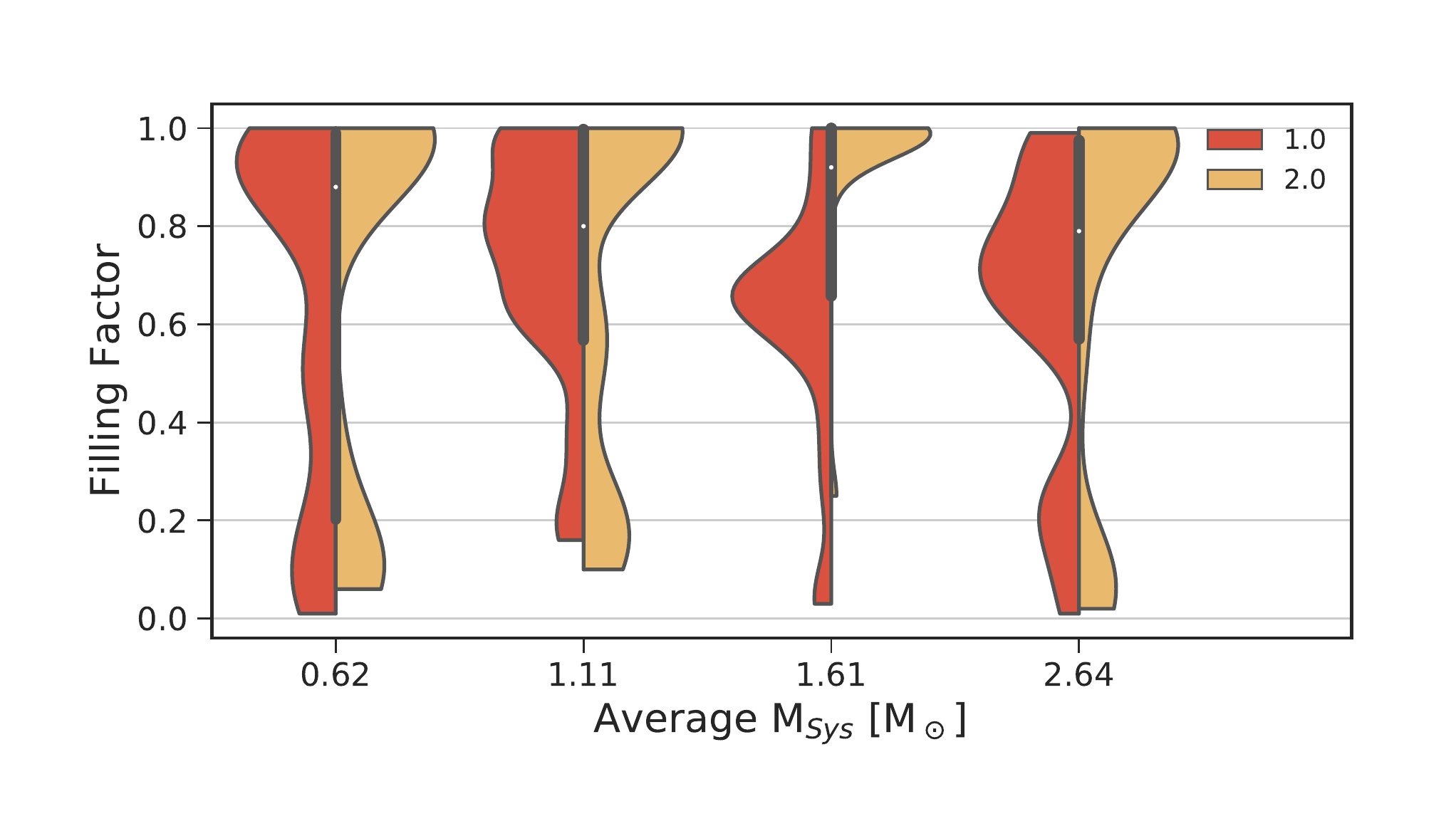}
\caption{Distribution of the filling factors for the primary (in red) and secondary (in orange) star in four different bins of masses. When the filling factor is equal to the unity, the star has filled its Roche lobe and is transferring mass to its companion.
}
\label{fig:binary_fill_fac}
\end{figure*}

The LCs of WTTS are thought to be dominated by the asymmetric distribution of cool or dark magnetic spots on the stellar surface, with stellar rotation being responsible for the modulation of the fluxes \citep{Bouvier1993, Herbst1994}. These sources show minimal to no evidence of ongoing accretion, possibly because the innermost part of their circumstellar disk have been already depleted.

Three percent of Wd2 stars fainter than $m_{F814W}=17.0$ show periodic (with periods ranging between 1.6 and 46 days, and with an average value of 7.7 days), small-amplitude  ($0.035 \times \Delta m_{F814W}\le 0.75$ mag) fluctuations consistent with Class {\sc iii} WTTSs. The position of the WTTS candidates in the $m_{F814}$ vs $m_{F814W}-m_{F160W}$ CMD is shown in {\it Panel A)} of Fig.~\ref{fig:variable_CMD2}. The number of WTTS candidates considerably decreases below $m_{F814W}\simeq 21.5$, with the faintest source found at $m_{F814W}\la 23.8$. This drop is likely caused by incompleteness effects. In a future paper, we will examine all the 47 orbits of HST time and perform artificial star tests to evaluate the impact of various selection biases, including temporal sampling, S/N and crowding, on the five populations of variables. {\it Panel A)} in Fig.~\ref{fig:variable_CC} shows that the WTTS candidates have relatively blue colors, in agreement with the hypothesis that these objects are likely quite evolved PMS stars.  

We used the  Stefan-Boltzmann law to convert the luminosities and temperatures derived from Padova isochrones into stellar radii ($1.5\le R_{WTTS}\le 5.7 R_\odot$). In doing this we found that  the rotational velocities $v sin(i)$ of the WTTS candidates range between 2.3 and $\sim 168$ km/s, with a peak at $\sim 12.5\, km/s$, in agreement with the $v sin(i)$ measurements found in the literature \citep[i.e.][]{Clarke2000, Herbst2002, Xing2006}.

\subsection{CTTS candidates}
\label{ctts}

CTTSs are class {\sc ii} young stellar objects that are still actively accreting material from their circumstellar disks. We identified 757 variable sources ($\sim 15\%$ of Wd2 low-mass stars) whose LCs are highly variable, but do not show evidence for periodicity on a short time-scale. These objects cover the magnitude range $17.0\la m_{F814W}\la 24.2$ ({\it Panel B)} of Fig.~\ref{fig:variable_CMD2}), and, compared to WTTS candidates, extend towards redder colors both in $m_{F814W}-m_{F125W}$ and in $m_{F814W}-m_{F160W}$  (Fig.~\ref{fig:variable_CC}). 

\subsection{Dipper candidates}
\label{dippers}

A special class of CTTS  has been discovered about 10 years ago. These objects are characterized by largely flat LCs, interrupted by sharp and short dips, and are therefore often reffered to as ``dippers'' \citep{Alencar2010, Morales-Calderon2011, Cody2014, Ansdell2016, Hedges2018}. Because the dips are normally shallower in the IR than at optical wavelengths, they have been attributed to large and dusty structures (such planetesimals) passing along our line of sight \citep{Bouvier2003, Alencar2010, McGinnis2015}, although their precise location within the disk \citep[inner vs. outer disk,][]{Bouvier1999, Bodman2017, Morales-Calderon2011} is still debated.  

\begin{figure}[ht!]
\centering
\includegraphics[width=8.4cm]{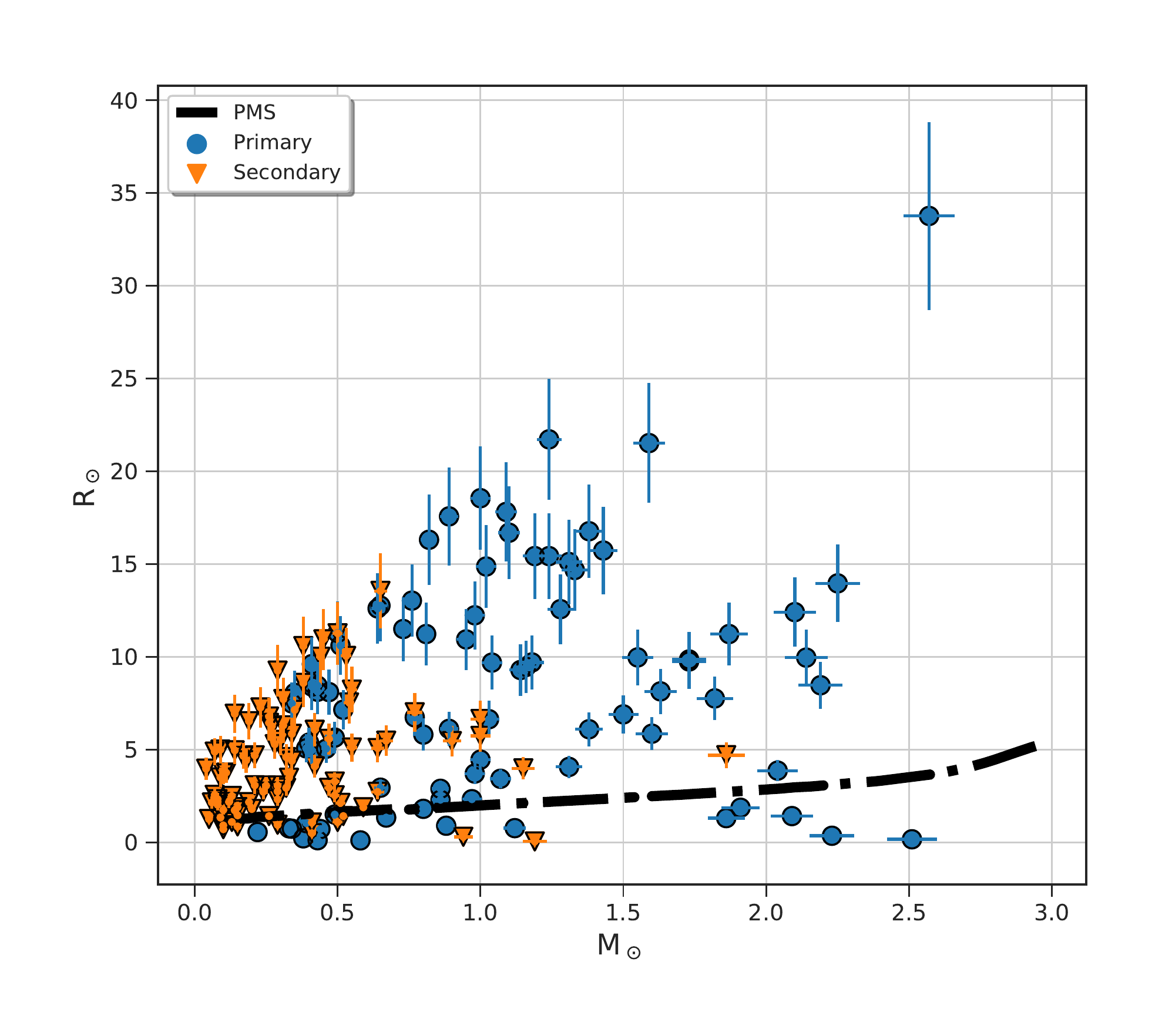}
\caption{Mass to radius relation for both the primary stars (shown as blue circles) and the secondary (orange triangles) as derived by fitting the phase-folded LCs of the EBs. The dotted-dashed black line shows the mass to radius relation for 1.6 Myr old PMS stars of similar masses and metallicites as derived from Padova isochrones.
}
\label{fig:m2r}
\end{figure}

72 of the sources in Wd2 between $19.14\le m_{F814W}\le 23.79$ ($0.3\le M \le 2.4 M_\odot$) show relatively flat LCs, interrupted by sudden drops in luminosity as large as to 2.78 magnitudes in the $m_{F814W}$ filter. Based on our temporal coverage, we conclude that the observed drops in luminosity are short events, that last at most few tens of days. An example of this type of objects is shown in Fig.~\ref{fig:mag_change}, where the luminosity of the star $\# 5320$ (RA$=10^h 24^m 05.463^s$,  Dec$=-57^o 45' 08.838''$, J2000) in the filter F814W dropped by $\sim 2.0$ magnitudes two times in less then two years.  

\subsection{Bursters}
\label{bursters}

In Wd2 about $8\%$ of the stars fainter than $m_{F814W}\la 16.62$ are characterized by eruptive behaviours, with the largest increase in magnitude we observed so far being $\Delta m_{F814W}=3.22$. In some cases the variation lasts for only few days, in others, once the star lights up, its luminosity remains elevated. 

Eruptive low-mass PMS stars are common and, based on the length of their burst, have been classified as either EX Orionis-type \citep[short burst][]{Herbig2007} or FU Orionis-type \citep[few year-long burst][]{Hartmann1996, Hartmann1998} stars. It is not clear yet if EXor and FUor-type stars belong to the same class of objects, nor if all the young low-mass stars undergo a bursting period. Nevertheless for both EXor and FUor-type stars, the large variation in brightness is ascribed to unstable pile-up of gas in the inner disk, which then releases a cascade of material onto the star. It is believed that most of the stellar mass is accumulated via these dramatic outbursts \citep[e.g.][]{Bell1994, Zhu2009, DAngelo2010}, during which a star can accrete up to $0.01 M_\odot$ of gas. 

{\it WISE} observations of  ``burster'' stars, identified during the K2 mission, show clear IR excess, typical of large inner circumstellar disks, and the strength of the outburst appears to correlate to the IR excess \citep{Cody2017}. These bursts are likely ``cooking'' the dust in the circumstellar disks, thus modifying their chemical composition \citep{Green2006, Quanz2007, Cieza2016}. Understanding the properties and the frequency of these bursting systems is therefore important for the models of stellar mass accretion, protoplanetary gas-rich disk evolution and, possibly, planet formation.

\subsection{Eclipsing binaries}
\label{ebs}

The phase-folded LCs of nearly $2\%$ of the objects between $0.1\la M \la 5.8$ $\rm{M_\odot}$ ({\it Panel E)} in Fig.~\ref{fig:variable_CMD2}) show two minima, characteristic of eclipsing binaries. For each of these objects we derived physical and orbital properties by iteratively fitting their phase-folded LCs with the publicly available software Nightfall.\footnote{See http://www.lsw.uni-heidelberg.de/~rwichman/Nightfall.html for the program and a user manual (Wichmann 1998).} 

We assumed the orbital period and the ephemeris zero-point derived by the Lomb-Scargle statistics, and the total mass inferred from Padova isochrones. Nightfall then compared the observed LC with simulations obtained for different mass ratios, stellar temperatures, inclination angles, and Roche lobe filling factors using a chi-square function to determine the best-fit solution. The derived values depend on the assumptions made in modeling the stellar atmospheres in the Nightfall software and in the Padova isochrones, used to infer the total masses of the binaries. Therefore they should be used only for comparisons and not as absolute values. Because of the sparse cadence and the relatively short temporal coverage of our observations, in this preliminary study, we assumed only circular orbits, and, in the case of interacting binaries, in the LC model we did not include the contribution of a mass-transferring disk. 

\begin{figure*}[ht!]
\centering
\includegraphics[width=15.cm]{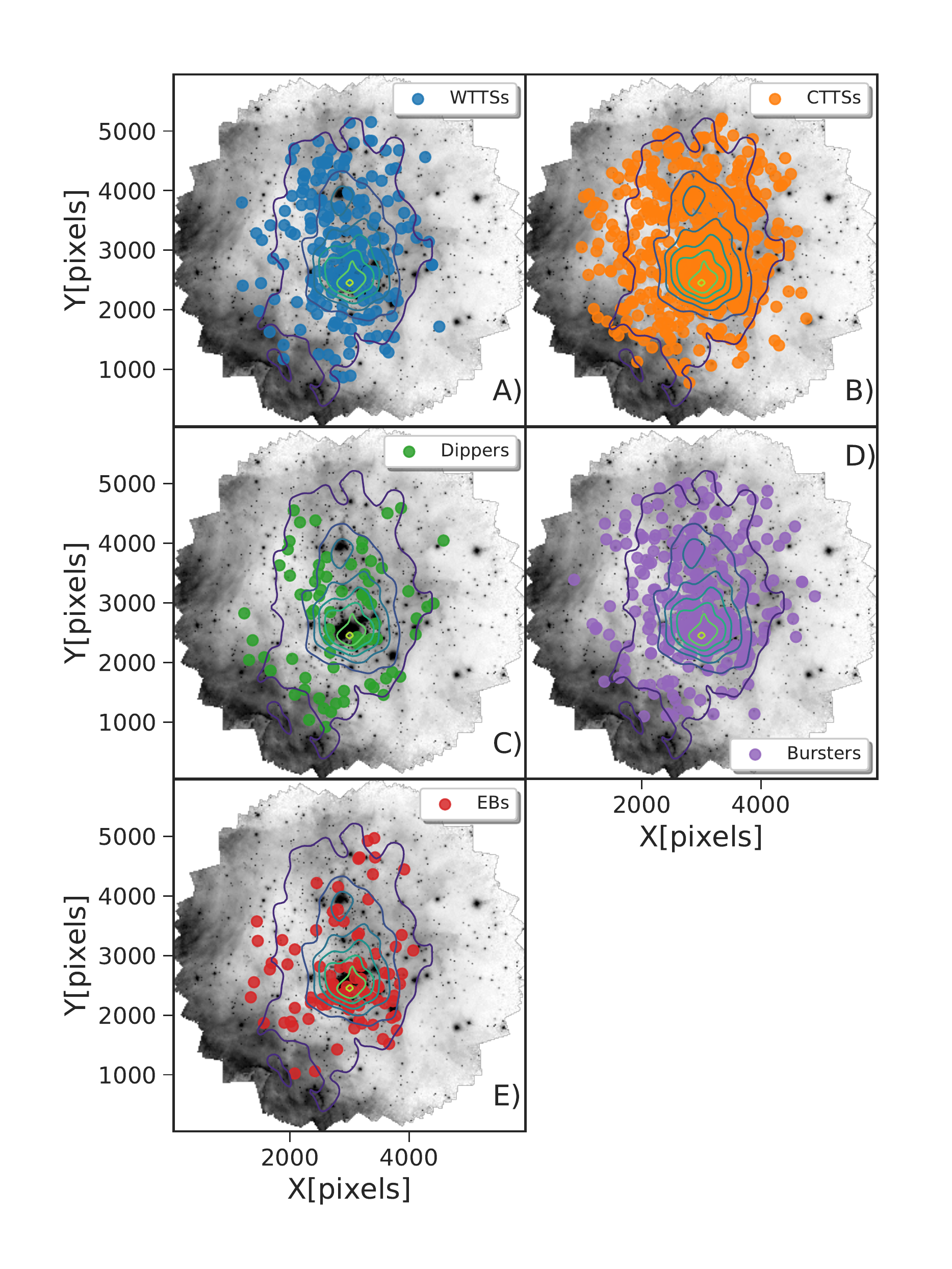}
\caption{Spatial distribution of Wd2 5 populations of variable stars superimposed on a black-and-white image of Wd2. Wd2 stellar isodensity contours are shown for reference. The populations are shown in five different panels following the same color-scheme used in Fig.~\ref{fig:variable_CMD2}.}
\label{fig:variable_map}
\end{figure*}

Because of the limited temporal sampling, our analysis is biased towards the shortest periods. Therefore we were able to find only eclipsing binaries with orbital periods ranging between $1.9\le P \le 86$ days, and separations between $6.8\le R \le 108\, R_\odot$ (0.03 - 0.5 AU). The {\it Upper Panel} of Fig.~\ref{fig:EB_properties} shows the distribution of periods covered in four different bins of system masses. The {\it Lower Panel} shows the distribution of mass ratios ($Q=M_2/M_1$) for the same bins of mass. It is interesting to note that in all four bins the distribution of the Q parameter is bimodal, with only $\sim 1/3$ of the binaries consisting of nearly equal-mass objects, similar to what observed by \citet{Raghavan2010} for binary systems, with primary star having mass $M_1\sim1\, M_\odot$. 

 \begin{figure*}[ht!]
\centering
\includegraphics[width=15cm]{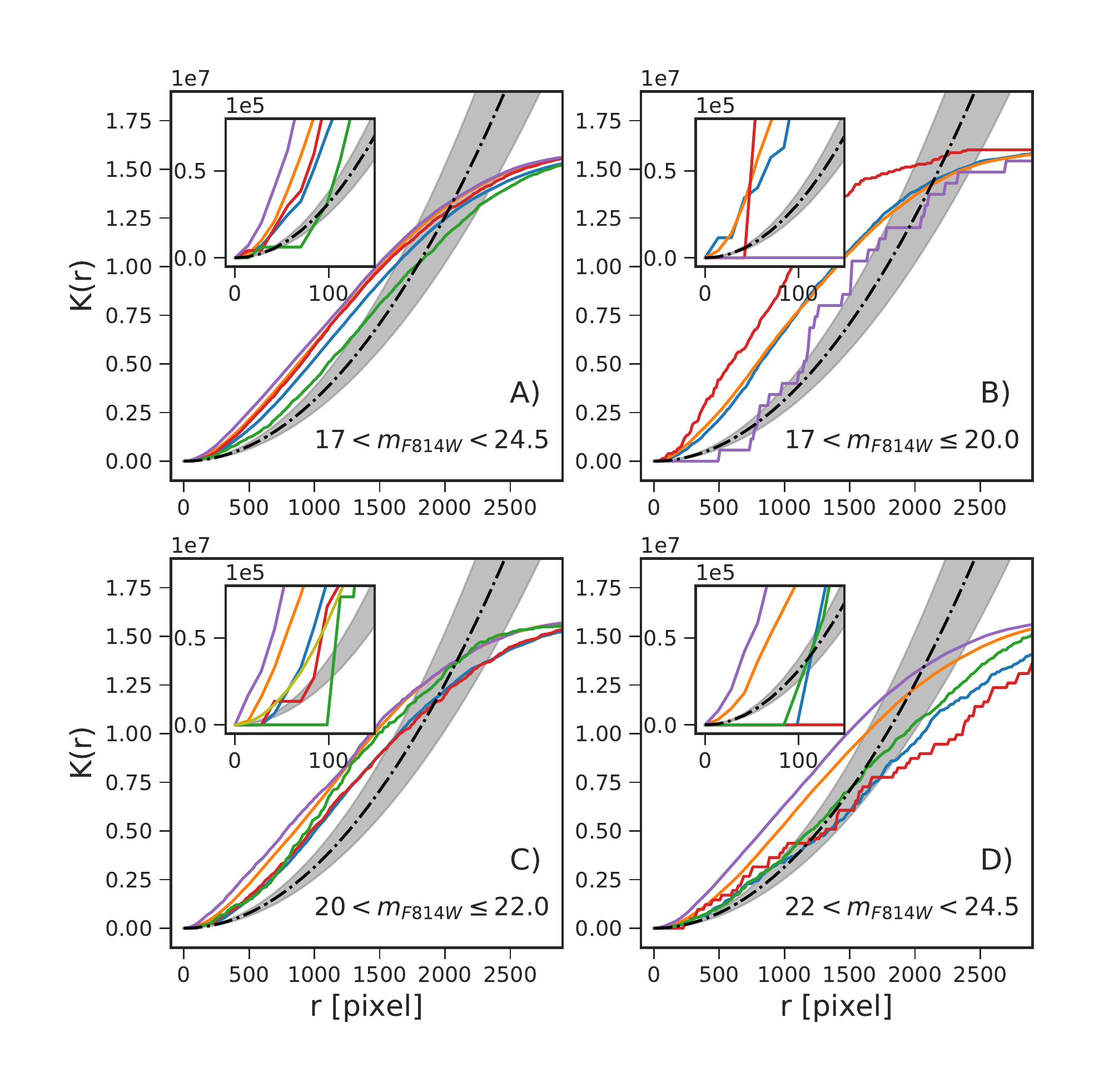}
\caption{Clustering properties, as derived from Ripley's K function, for for the WTTS candidates (blue line), CTTS candidates (orange line), dippers (green line), bursters (purple line) and EBs (red line). The response of the K function for a Poisson distribution and the 90\% confidence level are shown by the black dot-dashed line and the shaded area, respectively. {\it Panel A)} shows the clustering properties for all the variables fainter than $m_{F814W}>17.0$ (corresponding to $\sim 3.3\, M_\odot$ using Padova isochrones), {\it Panel B)} shows the behaviors of the stars between $17.0<m_{F814W}\leq 20.0$ (which correspond to the mass range between $\sim 1.75$ and $3.3\, M_\odot$, {\it Panel C)} refers to the stars between $20.0<m_{F814W}\leq 22.0$ ($\sim 0.7$ and $1.8\, M_\odot$), and Panel D) shows the properties of the fainter sources ($22.0<m_{F814W}< 24.5$, corresponding to the mass range $\sim 0.2$ and $0.7 M_\odot$ using Padova isochrones).}
\label{fig:K_test}
\end{figure*}

In 50\% of the systems the LCs are consistent with those of semi-detached binaries where one of the two components (the primary in 40\% of the cases and the secondary in the remaining 10\%) has filled its Roche lobe. In 2\% of the cases the LCs are consistent with those of contact binaries, with both stars having filled the Roche lobe. Because of the considerable fraction of interacting binaries, with ongoing mass transfer, it is likely that the mass ratio distribution will change over time, increasing the number of equal mass systems. The distribution of the filling factors for both the primary (in red) and the secondary star (in orange) for four different bins of masses is shown in Fig.~\ref{fig:binary_fill_fac}. 
 
 Fig.~\ref{fig:m2r} shows the mass to radius relation for the primary (blue circles) and the secondary (orange triangle) star as derived from the fitting of the EB LCs. The radii of the primary components range between $0.1 \le R_p \le 33.8 R_\odot$ and for the secondary between $0.1\le R_s \le 13.6$. In many cases these values are considerably larger than those derived from Padova isochrones using the Stephan-Boltzmann law. Phenomena such as tidal heating, ohmic dissipation and thermal tides have been considered to explain the inflated radii observed in close binaries and hot Jupiters, however in this case he fact that the library of models of atmosphere in Nightfall is not optimized to describe PMS stars and that the objects are quite far from the approximation of black-bodies is likely responsible for at least part of the discrepancies.

\subsection{Distribution of the variable PMS stars}

Fig.~\ref{fig:variable_map} shows the spatial distribution of the five population of variable PMS stars projected on image of Wd2 taken through the filter F814W. Isodensity contours for the stars belonging to Wd2 are shown for reference in all 5 panels. 

A visual inspection of Fig.~\ref{fig:variable_map} suggests that WTTS candidates ({\it Panel A)} are more concentrated around the two main clumps, while CTTS candidates ({\it Panel B)}) and bursters ({\it Panel D)} extend toward larger radii. Given that WTTS candidates are relatively massive objects, this is likely only a consequence of the fact that Wd2 is mass segregated \citep{Zeidler2015}.

In relatively low density star forming regions, like NGC~2264, dippers represent about 20-30\% of the CTTS population \citep{Alencar2010, Cody2017, Cody2018} and from the analysis of K2 mission data \citet{Hedges2018} found that dippers and bursters should have similar spatial distributions. However {\it Panel C)} shows that in Wd2 dippers clearly avoid the two regions of highest density, where the majority of the high-mass stars \citep[up to $80\, M_\odot$][]{Bonanos2004} are concentrated. In particular, based on the number of bursters and CTTS found within 150 pixels ($\sim 1.12$ pc) from the main clump of Wd2, in the same region we should find at least 5 dippers if they had the same spatial distribution of the other variable stars.

We used Ripley's K function \citep{Ripley1977, Ripley1979} to further investigate the clustering properties of the variable stars in Wd2. According to Ripley's statistics, a population is considered clustered within a certain scale $r$ if $K(r)$ is above the response of the K function for a Poisson distribution. On the contrary, distributions whose K function at the distance $r$ are below the Poisson distribution are considered dispersed. 
Figure~\ref{fig:K_test} compares the K functions for WTTS candidates (in blue), CTTS candidates (in orange), dippers (in green), bursters (in purple) and EBs (in red) to the K function of a Poisson distribution (black dashed line) over the entire range of magnitudes considered ({\it Panel A)}), for stars between $17<m_{F814W}\leq 20.0$ (corresponding to the mass range between 1.75 and 3.3 $M_\odot$ when using Padova isochrones, {\it Panel B)}), between $20<m_{F814W}\leq 22.0$ (corresponding to the mass range between 0.8 and 1.8 $M_\odot$, {\it Panel C)}), and between $22<m_{F814W}< 24.5$ (corresponding to the mass range between 0.2 and 0.8 $M_\odot$, {\it Panel D)}). 

Over the entire range of magnitudes ($17<m_{F814W}< 24.5$, Figure~\ref{fig:K_test} {\it Panel A)}) EBS, CTTS candidates and bursters have the same clustering scale $r=1.41\pm 0.14$ pc and their K functions  are indistinguishable from each other. On the same range of magnitudes, the clustering properties of WTTS candidates are less significant that those of the previous three populations, but their clustering scale is comparable ($r_{WTTS}=1.32 \pm 0.14$ pc).  

The clustering properties of the dippers between $17<m_{F814W}< 24.5$ are only marginally significant, and in the innermost $0.08\pm 0.01$ pc they are dispersed. Their clustering scale is also smaller than the other populations, being only $r_{Dip}=1.05 \pm 0.20$ pc. 

Because the cluster is mass segregated \citep{Zeidler2017}, we checked if Ripley's statistics could give differnet results over different ranges of magnitude/masses. If we consider only the variable PMS stars brighter than $m_{F814W}\leq 20.0$, we find that  EBs are significantly more clustered than CTTS and WTTS candidates.  In this range of magnitudes CTTS and WTTS candidates have identical distributions, while bursters are almost indistinguishable from a Poisson distribution. Only four dippers are brighter than $m_{F814W}\leq 20.0$, and therefore we do not considered their distribution in this range of masses.

Between $20.0<m_{F814W}\leq 22.0$, bursters have clustering properties and significance similar to CTTS candidates, while the properties of WTTS candidates resamble those of the EBs. On scales smaller than $r\simeq 800$ pixels ($\sim 0.7$ pc) dippers properties are similar to those of EBs and WTTS candidates , but at larger scales they become more similar to bursters and CTTS candidates. Finally for magnitudes fainter than $m_{F814W}> 22.0$ only CTTS candidates and bursters are significantly clustered.

The lack of dippers in the regions of highest stellar density suggests that the properties of stellar disks vary with the position in the cluster, and that in less than 2 Myr phenomena such as dynamical truncation \citep{Vinke2015, PortegiesZwart2016, Vincke2016}, stellar winds \citep{Pelupessy2012}, dynamical interaction \citep{Olczak2008, Reche2009, deJuanOvelar2012} and photoevaporation caused by the UV radiation coming from  nearby bright OB stars \citep[e.g.][]{Haworth2017}, can significantly affect disk growth and evolution. 

Similarly the different behaviour at different masses of the various type of variables can provide information on the life span and size of the circumstellar disks as a function of the mass of the star, although these trend will ahve to be further investigated to properly evaluate the effect of incompleteness.

\begin{figure}[ht!]
\centering
\includegraphics[width=8.4cm]{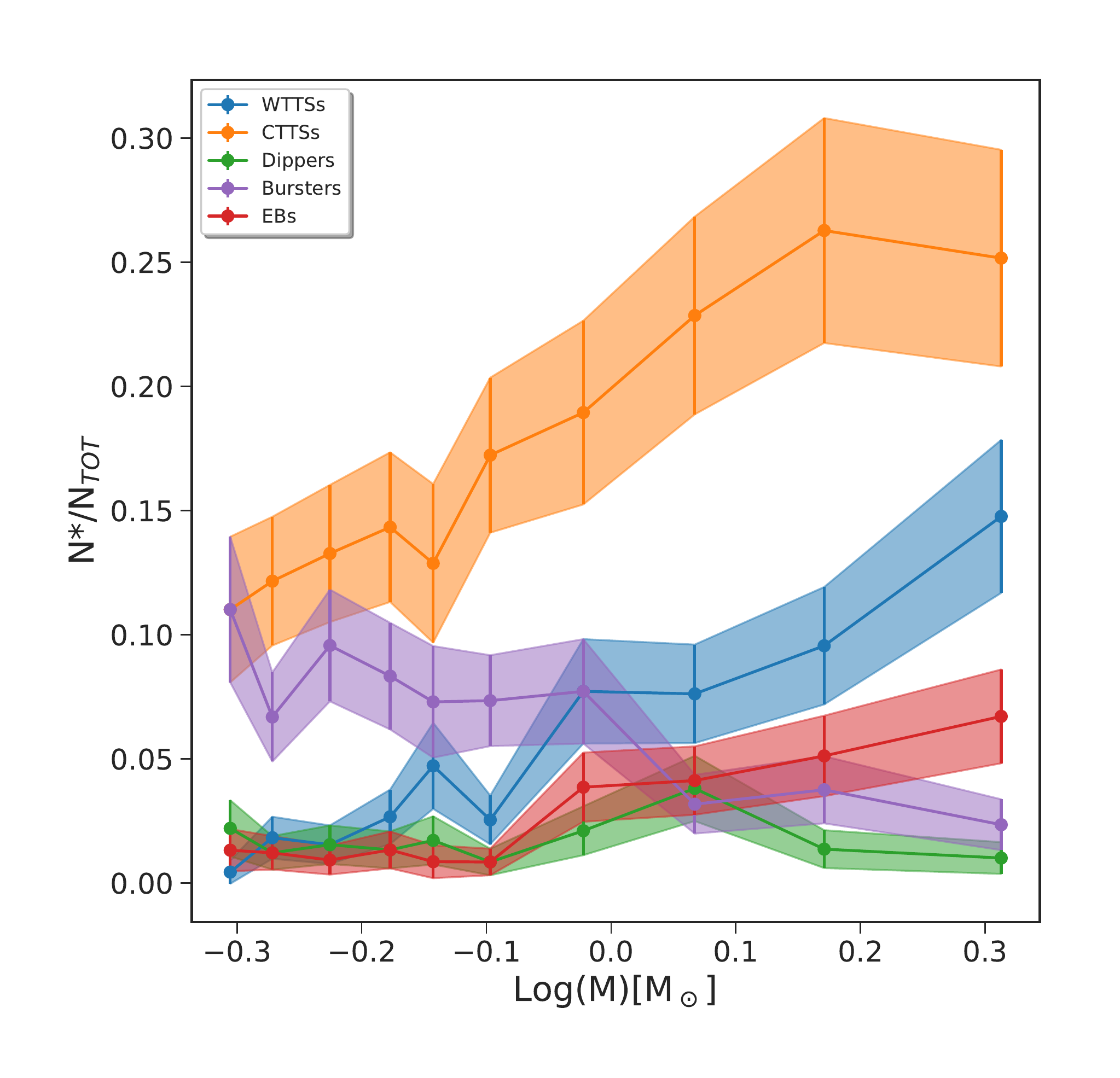}
\caption{Ratio between the number of variable stars and the total number of stars in Wd2 per bin of mass between 0.5 and 2.4 $M_\odot$. WTTS candidates are in blue, CTTS in orange, dippers in green, bursters in purple and EBs in red. 
}
\label{fig:pop_ratios}
\end{figure}

\section{Discussion and Conclusions}
\label{Conclusions}

The analysis of the Galactic YMC Wd2 with the WFC3 on board of HST showed that these systems are gold mines for studying the properties of variable PMS stars and investigate the evolution of their circumstellar disks as a function of mass and age. In fact at least 1/3 of the intermediate and low-mass PMS stars in Wd2 are variable. Based on the characteristics of their LCs, we classified the variable PMS stars as WTTS candidates, CTTS candidates, dippers and bursters. In addition, $2\%$ of the stars below $6 M_\odot$ are eclipsing binaries, with orbital periods shorter than 80 days.

By comparing the ratio of variable stars with respect to the entire cluster population per bin of mass  (Fig.~\ref{fig:pop_ratios}), we find that the fraction of both WTTS and CTTS candidates decreases as we move towards lower masses, and, in the case of WTTS candidates, we notice a clear change in slope  below $\sim 0.9\, M_\odot$. This drop in stellar counts is likely an artefact due to the fact that below $m_{F814W}\sim 22$ the LCs become too noisy to detect  periodic signals. In fact also the fraction of EBs  significantly decrease around this magnitude limit.

The bursters are the only group of variables to show a significantly different population ratio: their number remains almost constant between $\sim 0.5$ and $0.9\, M_\odot$ with the respect of other stars of similar masses and then considerably decreases at higher masses, suggesting that either this is a phenomenon mainly associated to relatively low mass stars, or that at higher masses the inner disks have been already significantly cleared and there is not enough material to feed strong bursts. In the latter cases Wd2 would provide us with a clock to estimate the evolution of the circumstellar disks as a function of stellar mass: while stars more massive than $\sim 1$ -- $2 M_\odot$ require less than $\sim 1.5$ Myr to considerably deplete their circumstellar disks, at lower masses the inner disks are likely more pristine, and can still provide a considerable amount of  material to be accreted by the star.

The comparison among the spatial distributions of the five populations of variables highlights how local conditions can affect the evolution of circumstellar disks. In particular, dippers appear to avoid the regions of higher stellar density, dominated by the high mass stars. Over the years several authors have investigated 
how dynamical truncation \citep{Vinke2015, PortegiesZwart2016, Vincke2016}, stellar winds \citep{Pelupessy2012}, dynamical interaction \citep{Olczak2008, Reche2009, deJuanOvelar2012} and photoevaporation from OB stars can affect the evolution of circumstellar disks. Recently, in simulating the effects of FUV radiation on circumstellar disks in clusters of different stellar densities, \citet{ConchaRamirez2019} estimated that in regions of high stellar density about 80\% of the disks can be destroyed by external photoevaporation in less then 2 Myr\, while, in comparison, mass loss caused by dynamical encounters is negligible. 

If the dramatic drop in luminosity experienced by the dippers is, as commonly accepted, due to the presence of large dusty structures and  planetesimals, the absence of dippers in the two higher density clumps of Wd2 could explain why planetary systems appear to be extremely rare in globular clusters \citep{gilliland00} and younger dense clusters \citep[e.g.][]{deJuanOvelar2012}. High spatial resolution followup observations in the near and mid-infrared using for example NIRSpec and MIRI on the James Webb Space Telescope will be needed to definitely characterize the properties of the disks in Wd2 and other YMCs to determine how local conditions affect the evolution of these systems and the formation of planetary systems.

In future papers we will extend our analysis to the entire data-set, and to the short exposures. With a better temporal sampling of the LCs, we will statistically estimate the duty cycles of the bursting episodes and better characterize the properties of the 5 populations of variables PMS stars as a function of mass. At the same time our better temporal coverage will allow us to explore a larger range of orbital periods for the population of EBs. Artificial star tests will allow us to quantify the impact of selection biases at different masses. Finally we plan to use image subtraction and proper motion analysis to identify and characterize the properties of long period binaries as a function of their positions in the cluster, and to investigate how dynamics and stellar feedback affect their formation and evolution.

\acknowledgments

We are grateful to the anonymous referee for the constructive suggestions that helped us to improve the quality of this paper. We thank Stefano Casertano, Guido De Marchi, Nino Panagia, and Mark Krumholz for the useful and constructive discussions. 
DJL acknowledges support from the Spanish Government Ministerio de Ciencia, Innovaci\'on y Universidades through grants PGC-2018-091\,3741-B-C22 and from the Canarian Agency for Research, Innovation and Information Society (ACIISI), of the Canary Islands Government, and the European Regional Development Fund (ERDF), under grant with reference ProID2017010115.

These observations are associated with programs \# 14087, 15362, 15514. Support for the programs \# 14087, 15362, 15514 was provided by NASA through a grant from the Space Telescope Science Institute. This work is based on observations obtained with the NASA/ESA Hubble Space Telescope, at the Space Telescope Science Institute, which is operated by the Association of Universities for Research in Astronomy, Inc., under NASA contract NAS 5-26555.

\acknowledgments
Facilities: HST(WFC3) - Hubble Space Telescope satellite, HST(ACS) - Hubble Space Telescope satellite.

\end{document}